\documentclass[]{article}
\usepackage{amssymb,latexsym,amsmath,enumerate,verbatim,amsfonts,amsthm}
\usepackage{cite}
\title{Note on ``Vacuum stability of a general scalar potential of a few fields"\footnote{This  work was supported by
		the National Natural Science Foundation of P.R. China (Grant No.12171064), by The team project of innovation leading talent in chongqing (No.CQYC20210309536) and by the Foundation of Chongqing Normal university (20XLB009)}}
\author{Yisheng Song\\
 \small  School of Mathematical Sciences, Chongqing Normal University, \\
\small Chongqing, P.R. China, 401331; Email: yisheng.song@cqnu.edu.cn}

\date{ }
\begin{document}

\maketitle

\begin{abstract}
This note is to give the analytic necessary and sufficient conditions  for the boundedness-from-below conditions of general scalar potentials of  two real scalar fields $\phi_1$ and $\phi_2$ and the Higgs bonson $\mathbf{H}$.

\end{abstract}

\section{Introduction}

Kannike \cite{K2016} presented the boundedness-from-below conditions of general scalar potentials of  two real scalar fields $\phi_1$ and $\phi_2$ and the Higgs bonson $\mathbf{H}$, \begin{equation}\label{eq:1}
	\begin{aligned}
		V(\phi_1,\phi_2,|H|)=& \lambda_{H}|H|^4+\lambda_{H20}|H|^2\phi_1^2+\lambda_{H11}|H|^2\phi_1\phi_2+\lambda_{H02}|H|^2\phi_2^2\\
		&\ +\lambda_{40}\phi_1^4+\lambda_{31}\phi_1^3\phi_2+\lambda_{22}\phi_1^2\phi_2^2+\lambda_{13}\phi_1\phi_2^3+\lambda_{04}\phi_2^4.
	\end{aligned}
\end{equation} This is equivalent to an analytic  necessary and sufficient conditions of $$V(\phi_1,\phi_2,|H|)>0\mbox{ for all }\phi_1, \phi_2, \mathbf{H}.$$

For two real scalar fields $\phi_1$ and $\phi_2$ and the Higgs bonson $\mathbf{H}$,  a  general scalar potentials $V(\phi_1,\phi_2,|H|)$ is  rewritten as follows$$
		V(\phi_1,\phi_2,|H|)=\lambda_{H}|H|^4+M^2(\phi_1,\phi_2)|H|^2+V(\phi_1,\phi_2),
$$ where
$$ M^2(\phi_1,\phi_2)=\lambda_{H20}\phi_1^2+\lambda_{H11}\phi_1\phi_2+\lambda_{H02}\phi_2^2,$$\begin{equation}\label{eq:1}  V(\phi_1,\phi_2)=\lambda_{40}\phi_1^4+\lambda_{31}\phi_1^3\phi_2+\lambda_{22}\phi_1^2\phi_2^2\\
	+\lambda_{13}\phi_1\phi_2^3+\lambda_{04}\phi_2^4.
\end{equation} So applying the  well-known positivity conditions of quadratic polynomial $$p(t)=at^2+bt+c$$ for all $t=|H|^2\geq0$ (which is showed hundreds of years ago),  $V(\phi_1,\phi_2,|H|)>0$ for all $\phi_1, \phi_2, \mathbf{H}$ ($a=\lambda_{H}>0$) if and only if  for all $\phi_1, \phi_2,$ 
\begin{equation}\label{eq:3}	\begin{cases}
		 c=V(\phi_1,\phi_2)>0, \mbox{ and either } b=M^2(\phi_1,\phi_2)\geq0; \mbox{ or }\\
		4ac-b^2=4\lambda_{H}V(\phi_1,\phi_2)-(M^2(\phi_1,\phi_2))^2>0.
\end{cases}\end{equation}
It is obvious that $M^2(\phi_1,\phi_2)=\lambda_{H20}\phi_1^2+\lambda_{H11}\phi_1\phi_2+\lambda_{H02}\phi_2^2$ is a quadric form with respect to two variables $\phi_1,\phi_2$, and hence, the inequality $M^2(\phi_1,\phi_2)\geq 0$ is equivalent to positive semi-definiteness of its coefficient matrix $M^2$. Then by Sylvester’s criterion, $M^2(\phi_1,\phi_2)\geq 0$  if and only if 
\begin{equation}\label{eq:4}
	\lambda_{H20}\geq0,\ \lambda_{H02}\geq0,\ \lambda_{H20}\lambda_{H02}-\frac14\lambda_{H11}^2\geq 0.
\end{equation}
So, Eqs.\eqref{eq:4} is  differ from Eqs.(54) and (55) of  Kannicke \cite{K2016}.  

\section{Boundedness-from-below conditions}
Now we correct this mistake and present  the analytic necessary and sufficient conditions are showed for the boundedness from below of scalar potential of two real scalar  fields $\phi_1$ and $\phi_2$ and the Higgs doublet $\mathbf{H}$.  It follows from the conclusion \eqref{eq:3} that we firstly need
  the analytic necessary and sufficient conditions of $V(\phi_1,\phi_2)>0$,
\begin{equation}\label{eq:1}  V(\phi_1,\phi_2)=\lambda_{40}\phi_1^4+\lambda_{31}\phi_1^3\phi_2+\lambda_{22}\phi_1^2\phi_2^2\\
	+\lambda_{13}\phi_1\phi_2^3+\lambda_{04}\phi_2^4.
\end{equation} It is obvious that the discriminant $D\geq0$ is a necessary  condition of $V(\phi_1,\phi_2)>0$.
 Such a  positivity condition may trace  back to ones of  Refs. Rees \cite{R1922} ,   Lazard \cite{L1988} Gadem-Li \cite{GL1964}, Ku \cite{K1965} and Jury-Mansour \cite{JM1981}.  Untill to 2005, Wang-Qi \cite{WQ2005} 
improved their proof, and perfectly gave  analytic necessary and sufficient conditions. For more  detail about  applications of these results,  see Song-Qi \cite{SQ2021} also.  That is, for all $\phi_1,\phi_2$ with $ (\phi_1,\phi_2)\ne(0,0)$,  the binary quartic homogeneous polynomial \eqref{eq:1}, 
$V(\phi_1,\phi_2)>0$ if and only if  
\begin{equation}\label{eq:2}\begin{cases}\lambda_{40}>0, \lambda_{04}>0,\\
		D=0, \ G=0,\ R=0 \mbox{ and } Q>0; \\
		D>0\mbox{ and } Q\geq0, \mbox{ or } R>0
\end{cases}\end{equation}
where $$\begin{aligned}
	G=& \frac14\lambda_{40}^2\lambda_{13}-\frac1{8}\lambda_{40}\lambda_{31}\lambda_{22}+\frac1{32}\lambda_{31}^3\\
	Q=& \frac16\lambda_{40}\lambda_{22}-\frac1{16}\lambda_{31}^2=\frac1{48}(8\lambda_{40}\lambda_{22}-3\lambda_{31}^2)\\
	I=&\lambda_{40}\lambda_{04}-\frac1{4}\lambda_{31}\lambda_{13}+\frac1{12}\lambda_{22}^2\\
	J=&\frac16\lambda_{40}\lambda_{22}\lambda_{04}+\frac1{48}\lambda_{31}\lambda_{22}\lambda_{13}-\frac1{216}\lambda_{22}^3\\&-\frac1{16}\lambda_{40}\lambda_{13}^2-\frac1{16}\lambda_{31}^2\lambda_{04}\\
	D=&I^3-27J^2, R=\lambda_{40}^2I-12Q^2.
\end{aligned}$$
Recently, Qi-Song-Zhang \cite{QSZ2022} gave a new  necessary and sufficient condition other than the above results \eqref{eq:2} in forms. \\ 

Next  we  give the revised version of the conclusion Eq.(68) in Kannicke \cite{K2016}. Let $V'(\phi_1,\phi_2)=4\lambda_{H}V(\phi_1,\phi_2)-(M^2(\phi_1,\phi_2))^2$. Now we show $V'(\phi_1,\phi_2)>0$. 
$$\begin{aligned}
		V'(\phi_1,\phi_2)=&4\lambda_{H}V(\phi_1,\phi_2)-(M^2(\phi_1,\phi_2))^2\\=&(4\lambda_{40}\lambda_H-\lambda_{H20}^2)\phi_1^4+(4\lambda_H\lambda_{31}-2\lambda_{H20}\lambda_{H11})\phi_1^3\phi_2\\
	&+(4\lambda_H\lambda_{22}-2\lambda_{H20}\lambda_{H02}-\lambda_{H11}^2)\phi_1^2\phi_2^2\\
	&+(4\lambda_H\lambda_{13}-2\lambda_{H02}\lambda_{H11})\phi_1\phi_2^3+(4\lambda_{04}\lambda_H-\lambda_{H02}^2) \phi_2^4\\
	=&\lambda_{40}'\phi_1^4+\lambda_{31}'\phi_1^3\phi_2+\lambda_{22}'\phi_1^2\phi_2^2+\lambda_{13}'\phi_1\phi_2^3
	+\lambda_{04}'\phi_2^4,
\end{aligned}$$
where
$$
\begin{aligned}\lambda_{40}'&=4\lambda_{40}\lambda_H-\lambda_{H20}^2,\  \lambda_{04}'=4\lambda_{04}\lambda_H-\lambda_{H02}^2,\\
	\lambda_{31}'&=4\lambda_H\lambda_{31}-2\lambda_{H20}\lambda_{H11},\ \lambda_{13}'=4\lambda_H\lambda_{13}
	-2\lambda_{H02}\lambda_{H11},\\
	\lambda_{22}'&=4\lambda_H\lambda_{22}-2\lambda_{H20}\lambda_{H02}-\lambda_{H11}^2.
\end{aligned}$$
In terms of the coefficients of $V'(\phi_1,\phi_2)$, we define  the following quantities: 
$$\begin{aligned}
	G'=& \frac14\lambda_{40}'^2\lambda_{13}'-\frac1{8}\lambda_{40}'\lambda_{31}'\lambda_{22}'+\frac1{32}\lambda_{31}'^3\\
	Q'=& \frac16\lambda_{40}'\lambda_{22}'-\frac1{16}\lambda_{31}'^2\\
	I'=&\lambda_{40}'\lambda_{04}'-\frac1{4}\lambda_{31}'\lambda_{13}'+\frac1{12}\lambda_{22}'^2\\
	J'=&\frac16\lambda_{40}'\lambda_{22}'\lambda_{04}'+\frac1{48}\lambda_{31}'\lambda_{22}'\lambda_{13}'-\frac1{216}\lambda_{22}'^3\\&-\frac1{16}\lambda_{40}'\lambda_{13}'^2-\frac1{16}\lambda_{31}'^2\lambda_{04}'\\
	D'=&I'^3-27J'^2, R'=\lambda_{40}'^2I'-12Q'^2.
\end{aligned}$$
Then an application of  the conclusion \eqref{eq:2}, we have $V'(\phi_1,\phi_2)>0$ for all $\phi_1,\phi_2$ with $ (\phi_1,\phi_2)\ne(0,0)$ if and only if  
\begin{equation}\label{eq:6}\begin{cases}\lambda'_{40}>0, \lambda'_{04}>0,\\
		D'=0, \ G'=0,\ R'=0 \mbox{ and } Q'>0;  \\
		D'>0\mbox{ and } Q'\geq0 \mbox{ or }  R'>0.
\end{cases}\end{equation}

Altogether, combing  Eq. \eqref{eq:3} and Eqs. \eqref{eq:2}, \eqref{eq:4},  \eqref{eq:6}, the analytic  necessary and sufficient condition is established for the boundedness from below of scalar potential of two real scalar  fields $\phi_1$ and $\phi_2$ and the Higgs doublet $\mathbf{H}$ . That is, 
$V(\phi_1,\phi_2,|\mathbf{H}|)>0$ for all $\phi_1, \phi_2, \mathbf{H}$ with $ (\phi_1,\phi_2, \mathbf{H})\ne(0,0,0)$ if and only if  
\begin{equation}\label{eq:7}\begin{cases} \lambda_{H}>0, \lambda_{40}>0, \lambda_{04}>0\mbox{ and }\\
		(i)	\  \  \ D=0, \ G=0,\ R=0 \mbox{ and } Q>0;  \\
		\  \ \ \ \ D>0\mbox{ and either } Q\geq0, \mbox{ or } R>0; \mbox{ and either}\\
		(ii)\ \ \lambda_{H20}\geq0,\ \lambda_{H02}\geq0,\ 4\lambda_{H20}\lambda_{H02}-\lambda_{H11}^2\geq 0;\mbox{ or }\\
		 (iii)\  \  4\lambda_{40}\lambda_H-\lambda_{H20}^2>0,\ 4\lambda_{04}\lambda_H-\lambda_{H02}^2>0\mbox{ and}\\
		\  \ \ \ \ D'=0, \ G'=0,\ R'=0 \mbox{ and } Q'>0; \\
		\  \ \ \ \ D'>0\mbox{ and  } Q'\geq0, \mbox{ or }  R'>0.
\end{cases}\end{equation}

%\section*{Acknowledgments} This  work is supported by the National Natural Science Foundation of P.R. China (Grant No. 12171064) and by the Foundation of Chongqing Normal university (20XLB009).%The authors would like to express their sincere thanks to the anonymous referees for his/her constructive comments and valuable suggestions.
% ----------------------------------------------------------------
%\bibliographystyle{amsplain}

\end{document}